\documentclass[11pt,twoside]{article}

\usepackage{asp2006}
\usepackage{epsf}

\begin{document}
\title{Extinction curves in AGN}   
\author{B. Czerny}   
\affil{Copernicus Astronomical Center, Warsaw, Poland}    

\begin{abstract} 
The presence of the dust in the circumnuclear region strongly affects our
view of the nucleus itself. The effect is strong in type 2 objects but weaker
effect is likely to be present in type 1 objects as well. In these objects
a correction to the observed optical/UV spectrum must be done in order to 
recover the intrinsic spectrum of a nucleus. The approach based on the 
extinction curve is convenient for that purpose so significant effort has been
recently done in order to determine the extinction curve for the circumnuclear
material. It seems clear that the circumnuclear dust is different from the
average properties of the dust in the Interstellar Medium in our galaxy: the
well known 2175 \AA~ feature is weak or absent in AGN nuclear dust, and
the extinction curve at shorter wavelength does not seem to be rising as 
steeply. The circumnuclear dust is therefore more similar to SMC dust, or 
more likely, to the dust in very dense molecular clouds in our Galaxy. 
However, the
exact shape of the extinction curve in the far UV is still a matter of debate,
and various effects are difficult to disentangle.
\end{abstract}



\section{Introduction}

The dust present in the circumnuclear region shapes the spectra of active nuclei. It intercepts a significant fraction of the radiation generated by the matter accreting onto the central black hole. The effect is seen in the optical/UV/soft X-ray band as extinction, and in the IR band as emission. Since most of the accreting matter is concentrated in the vicinity of the equatorial plane the largest amount of dust seems to be concentrated there. The sources seen at large inclinations with respect to the symmetry axis are most affected by dust and they are recognized as type 2 objects (Seyfert 2 galaxies, type 2 QSO, Narrow Line Radio Galaxies) in which we have so direct view to the nucleus. In type 1 objects (Seyfert 1 galaxies, QSO, Broad Line Radio Galaxies) we see the nucleus directly. This picture was recognized by Antonucci \& Miller (1985) and it is now generally adopted under the name of the standard unification scheme (for recent reviews, see e.g. Veron-Cetty \& Veron 2000, Elitzur 2006). Although nowadays there are many arguments that the static geometrically thick dusty/molecular torus postulated by Antonucci \& Miller (1985) should be replaced with dynamical clumpy wind from an underlying disk (e.g. Jaffe et al. 2004, Elitzur \& Shlosman 2006) the basic scheme remains the same.

The unification picture means that if the study concentrates on type 1 objects the dust effect in the optical/UV band cannot be too large. However, it does not mean that the line of sight towards the nucleus is dust-free. Certain amount of dust is likely to be there and it may affects the observed spectrum, particularly in the UV band. Any study of the intrinsic emission of an active nucleus must take that into account. Since the optical depth of the dust is not large the intrinsic spectrum can be recovered from the observed spectrum if we know the wavelength-dependent extinction curve and the total amount of dust along the line of sight. This is much simpler than solving geometry-dependent 3-D radiative transfer within the dust. However, it does not mean that it is simple since the shape of the extinction curve representative for the dust in the circumnuclear region of an active nucleus is still a subject under discussion.

\section{Basic remarks on extinction curve approach}

The notion of extinction was introduced for stars in our Galaxy in the context of their location in the color-color diagram so the notation used is still strongly related to those traditional concepts of optical astronomy. Extinction at a specific wavelength $\lambda$ is defined as 
\begin{equation}
A_{\lambda} = m_{\lambda} -  m_{0,\lambda},
\end{equation}
i.e. as a difference between the observed magnitude of an object and the magnitude the object would have in the absence of the dust. Since the emitted flux is diluted by the dust roughly as $F_{obs}=F_{int}exp(-\tau_{\lambda})$, this means that the extinction is proportional to the dust optical depth. If we want now to characterize the wavelength dependence we must introduce the quantity which would be roughly independent from the density column of the dust. There are a few different such definitions used in the literature. For example, we can use the extinction ratio, $A_{\lambda}/A_{I}$, where $ A_{I}$ is the extinction of the same source at some fixed IR wavelength (see e.g. Draine 2003a). However, traditionally two other popular version of the normalization are used
\begin{equation}
X(\lambda) = {A_{\lambda} \over E(B-V)},~~~~~~~~~{\rm where ~~} E(B-V)=A_B-A_V
\end{equation} 
or
\begin{equation}
X(\lambda) = {A_{\lambda} - A_V \over E(B-V)}.
\end{equation}
The excess color $E(B-V)$ is also used to define the slope of the extinction curve, $R_V = A_V/E(B-V)$. The second definition is used more frequently since its determination does not require the absolute calibration of the stellar fluxes. However, in this case the quantity $R_V$ should be supplied independently. The amount of reddening of a given star is then usually given as the $E(B-V)$ value, which can be used to estimate the extinction in the visual band, $A_V$, taking the typical value of $R_V = 3.1$ for the diffuse interstellar medium in our Galaxy (Schultz \& Wiemer 1975), and we can get the total hydrogen column of the material along the line of sight from the formula $N_H = (1.79 \pm 0.03) A_V \times 10^{21}$ cm$^{-2}$, derived from the ROSAT study (Predehl \& Schmidt 1995).

\section{Standard extinction curve in our Galaxy and 2175 \AA~ feature}

The extinction curve is determined observationally, by comparing the spectrophotometry of two objects: one with negligible dust effect and one which is heavily reddened. However, the wavelength-dependent ratio of the two spectra give the extinction curve correctly only if we use the {\it two stars of the same spectral class} which ensures that the two spectra are intrinsically the same. Finding such pairs of stars is relatively easy in the case of stars in our Galaxy. However, finding an appropriate pair is a major problem when we deal with active galactic nuclei, and we will return to this issue in Sect.~\ref{sect:critical}.  

The extinction due to dust in the Interstellar Medium in our Galaxy was first determined in the optical band and it was found to be rising up as $\lambda^{-1}$ (Trumpler 1930, Greenstein 1938). Later research could cover UV part of the spectrum due to rocket flights (Byram et al. 1957). The extinction generally rises towards UV but it shows a peculiar feature around 2200 \AA, first found by Stecher (1965) and confirmed by later studies (e.g. Fitzpatrick \& Massa 1990, Valencic et al. 2004). The feature was immediately attributed to the presence of graphite in the interstellar medium (Stecher \& Donn 1965). The actual nature of this feature is still under discussion (see e.g. Draine 2003a, Duley \& Lazarev 2004) but some natural extensions of the initial hypothesis (e.g. {\it Policyclic Aromatic Hydrocarbon} - PAH – molecules, Wada et al. 1999) are particularly attractive. The whole extinction curve was nicely parameterized by Seaton (1979), for more recent models see Witt \& Gordon (2000) and Draine (2003b).

\begin{figure}
\epsfxsize = 80 mm \epsfbox{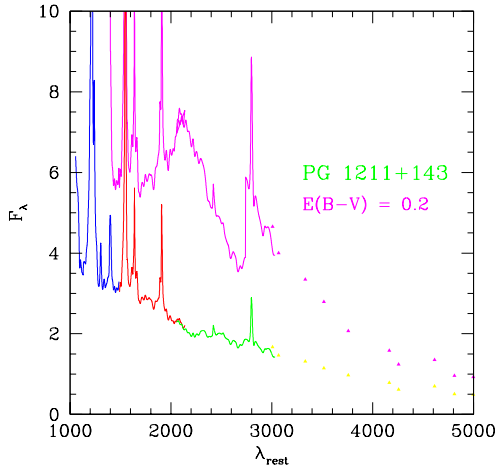}
\caption{The broad band spectrum of the quasar PG1211+143 (lower curve), and a result of derredening of this spectrum using the Seaton extinction curve. The corrected spectrum shows a strange bump caused by the 2175 \AA~ feature in Seaton curve.}
\label{fig:pg_seaton}
\end{figure}

Such an extinction curve, if used to recover the intrinsic spectrum of a quasar, leads to rather peculiar result, as illustrated in Fig.~\ref{fig:pg_seaton}. Therefore, in the early days of AGN research, the lack of 2175 \AA~ feature in quasar spectra was used as an argument against the presence of the dust in these objects (e.g. McKee \& Pertosian 1974, Mushotzky 1984).

\section{Extinction curves at various astrophysical sites}

Further research showed, however, that the dust properties captured by the Seaton extinction curve are not universal. The extinction towards the stars in 30 Doradus region in LMC shows much weaker 2175 \AA~ feature than the Milky Way dust, although the dust away from 30 Doradus is more similar to MW dust. The dust in the bar region of SMC is almost completely devoid of 2175 \AA~ feature (Prevot et al. 1984). The strong dependence of the shape of the extinction curve on the specific line of sight in LMC and SMC is clearly visible in the recent {\it Far Ultraviolet Spectroscopic Explorer} (FUSE) results (Hutchings \& Giasson 2001, Cartledge et al. 2005).

In Milky Way there are also some differences between various lines of sight, mostly seen factor $R_V$ (extreme values of 2.1 and $\sim 5.7$ were found, see Draine 2003a for references) but in most cases the differences are not large. Exceptions are connected  with dense molecular regions, like for example Taurus dark cloud. The extinction by the dense clump TMC-1 shows weak or absent 2175 \AA~ feature (Whittet et al. 2004).

The extinction law for starburst galaxies was studied by Kinnney et al. (1994) and Calcetti et al. (1994). The extinction curve was found to be featureless, like in SMC, but rising not so steeply as in SMC towards shorter wavelength. Distant galaxies, at redshifts between 1 and 2, studied through the analysis of the dust in the intervening systems revealed the presence of the SMC-type dust (York et al. 2006), and the similar result was obtained for lensing galaxies by Eliasdottir et al. (2006). There are  some detections of the 2175 \AA~ feature in some distant absorption systems or in distant galaxies (e.g. Cohen et al. 1999, see Wang et al. 2004 for other references) but the detections are frequently uncertain. Noll et al. (2006) argue that that a subclass of luminous starburst galaxies with 2175 \AA~ feature does exist but at limited range of redshifts.  The search in SDSS for quasars with 2175 \AA~ feature due to intervening systems was done by Wang et al. (2004). Only several candidates were found. Three most convincing candidates were analysed in detail. The best-fit reddening parameters were $E(B-V)$ from 0.13 to 0.23, and $R_V$ from 0.7 to 5.5, thus indicating wide range of grain properties among intervening systems.

This means that the dust in AGN nuclei is rather unlikely to have similar properties to MW dust.

\section{AGN extinction}

The 'non-standard' dust properties in AGN were first noticed during the attempts of the modelling the dust emission in the IR. Modification of the grain composition (Czerny et al. 1992, Loska et al. 1993) as well as grain sizes (Laor \& Draine 1993) in comparison to the MW dust seemed unavoidable. Recent results confirmed these results (e.g. Maiolino et al. 2001a,b, Siebenmorgen et al. 2005, Sturm et al. 2005). Theoretical extinction curves based on the modified dust properties showed little or no 2175 \AA~ feature and seemed also appropriate for dereddening of the UV spectra of NGC 6814 (Czerny et al. 1995). Similar extinction curves were determined observationally for NGC 3227 (Crenshaw et al. 2001) and Ark 564 (Crenshaw et al. 2002) by comparing their spectra to other bluer AGN. Extinction curve for a dust in a very distant quasar (SDSSJ1049+46, z=6.2) obtained by Maiolino et al. (2004) is again featureless, but at the same time it is significantly different from SMC curve.

Further significant progress in the field was possible when large samples of AGN spectra became available.  

\subsection{Results based on composite spectra} 

Gaskell et al. (2004) used the Molonglo Quasar Sample (MQS) for 72 radio loud AGN and the Large Bright Quasar Survey (LBQS) of 1018 for mostly radio quiet objects. They derived the extinction curve for radio loud and radio quiet objects independently. In the first case they gruped objects into subsamples according to the object orientation, in the second case they compared the composite of Francis et al. (1992) with the relatively unreddened composite of Baker \& Hunstead (1995). The derived extinction curve for radio quiet objects was somewhat steeper then the corresponding curve for radio loud objects but both were flatter, or more grey, than the standard MW curve, and the 2175 \AA~ feature was largely absent.

Similar work was done independently by Czerny et al. (2004) at the basis of the composite spectra from Sloan Digital Sky Survey (SDSS) obtained by Richards et al. (2003). In this case the bluest composite was used as a reference. The resulting curve is slightly steeper than in Gaskell et al. for radio quiet objects but it is also featureless, nevertheless it was not as steep as the extinction curve in SMC. The observationally derived curve was satisfactory fitted by the amorphous carbon dust with rather large grain sizes, above 0.016 $\mu$m. 

The approach based on the composits was however criticized by Willot (2005) who pointed out that composites are made of different objects at different wavelengths which may introduce significant bias. He argued that if the correction for this bias is made, the extinction curve is actually as in SMC.

\begin{figure}
\epsfxsize = 80 mm \epsfbox{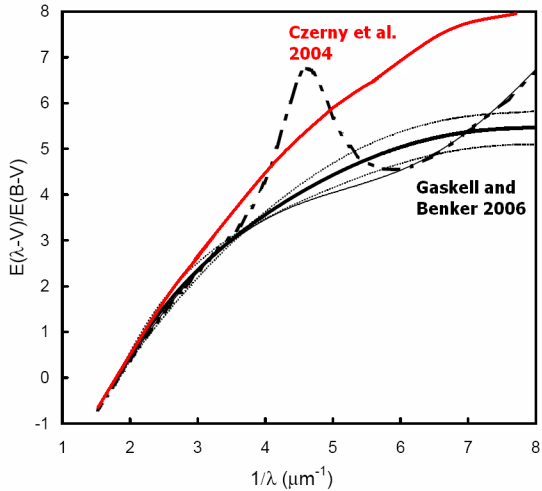}
\caption{The extinction curve of Czerny et al. (2004) and of Gaskell and Benker (2006) for radio quiet AGN. Dashed curve shows the Seaton curve.}
\label{fig:czerny}
\end{figure}

\subsection{Results based on large number of individual objects}

First such extensive study based again on SDSS sample was performed by Hopkins et al. (2004). They parameterized the slopes and the curvatures of the individual spectra and compared them to the mean curvatures determined for each redshift bin separately. The resulting curvature vs. slope plot was compared to predictions made at the basis of the available extinction curves. The distribution was best match by the predictions based on SMC curve, thus going in line with the result of Willot (2005).

On the other hand, Gaskell \& Benker (2006) derived 14 extinction curves for individual data and argue that only one of these extinction curves is as steep as the one in SMC. Other curves are flatter, and some actually show the presence of 2175 \AA~ feature, in one case the feature is as strong as in MW dust. The data used in this project come from Shang et al. (2005) and consists of observations performed by FUSE and {\it Hubble Space Telescope} (HST). Three objects from this sample were used as a reference. The intrinsic extinction derived in this project was the highest for PG1351+640 ($E(B-V)=0.365$), the extinction curve derived for this object is likely to be most reliable. This curve is featureless, not as steep as SMC, flatter than the curve in Czerny et al. (2004) but steeper that the Gaskell et al. (2004) result. The average extinction curve derived in this project is shown in Fig.~\ref{fig:czerny}. It is similar to the curve for PG1351+640 alone, supporting the flattening of the extinction in far UV unlike the behaviour of SMC curve. 

\subsection{Critical remarks}
\label{sect:critical}

\begin{figure}
\parbox{\textwidth}{
\parbox{0.5\textwidth}{
\epsfxsize=0.49\textwidth
\epsfbox[50 200 540 720]{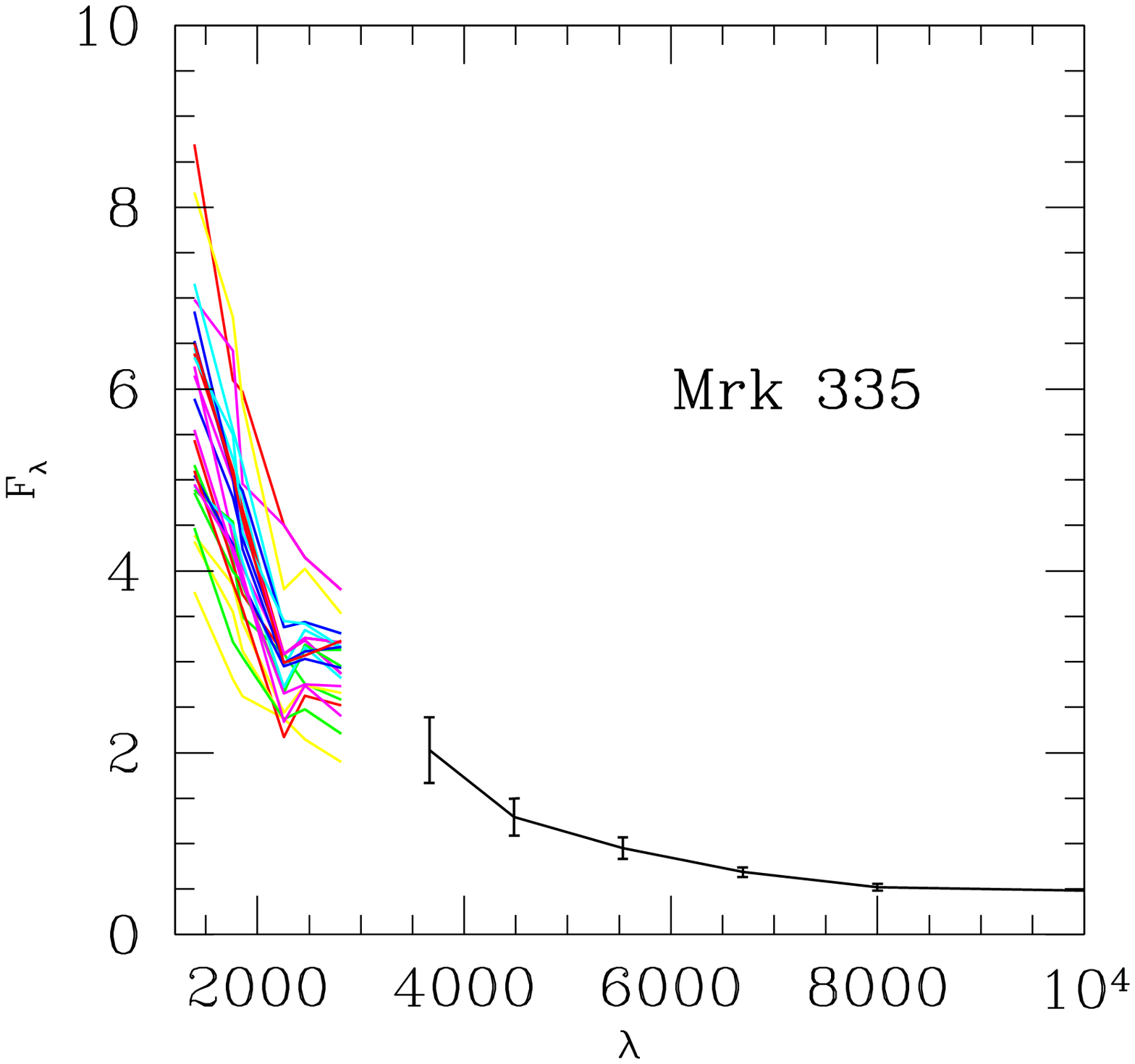}}
\hfil
\parbox{0.5\textwidth}{
\epsfxsize=0.49\textwidth
\epsfbox[50 200 540 720]{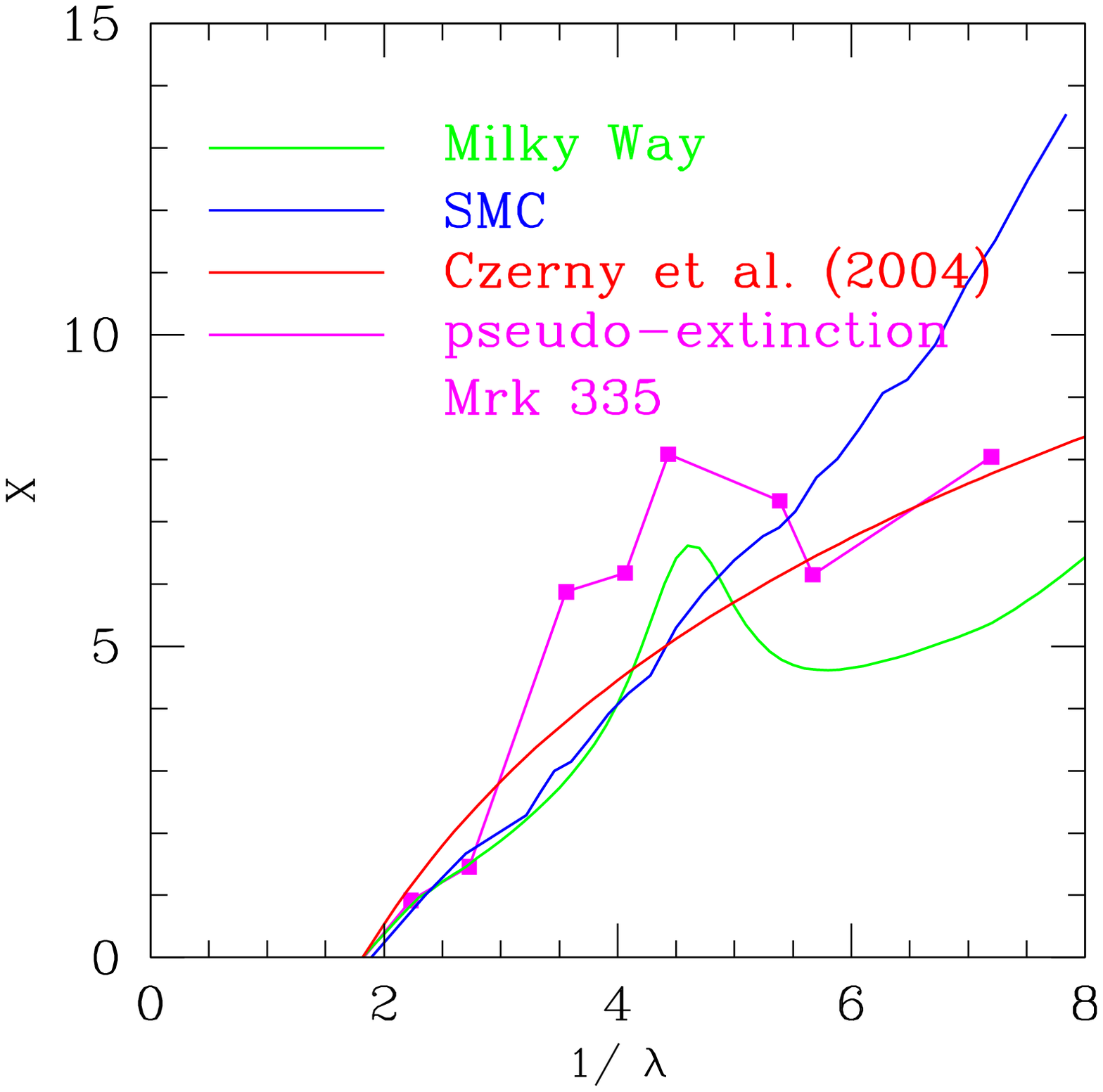}}
}
\caption{Left panel: the set of IUE spectra of Mkr 335 from Dunn et al. (2006) supplemented with the optical spectrum of Doroshenko et al. (2005), with optical error bars giving 1 $\sigma$ variability amplitude. Right panel: the comparison of the pseudo-extinction curve determined from the left panel data with other extinction curves. See Sect.~\ref{sect:critical} for the discussion.}
\label{fig:mrk335}
\end{figure}

The derivation of the extinction curve is complicated by the fact we have no guarantee that the pair of objects differ only with repect to the reddening. In stellar astrophysics, we have a support from stellar classification system. In AGN we also have some classes but they are not equally precise. What is worse, they are unlikely to be ever so precise since stars are roughly spherically symmetric while AGN are not. The AGN classification is always likely to be connected somehow to the inclination angle but the amount and the properties of dust are also likely to depend on the inclination. Therefore, reddened and unreddened objects are always quite likely to differ either with respect to the inclination or with respect to the luminosity which means that their intrinsic spectra are not going to the same.

The possible bias can be tackled by experimenting with the data or with the models. Gaskell \& Benker (2006) used a power law and a single black body component to mimick the AGN spectra and derived pseudo-extinction curves by comparing the two spectra with different value of the black body component. They also experimented with two power laws, and with different amount of the starlight contribution. Obtained pseudo-extinction curves cover a broad range of possibilities, some of them not unlike SMC curve, although the model tends to impose a coupling between the derived shape of the pseudo-extinction curve and the maximum reddening possible. The authors conclude that the shapes consistent with observed spectra do not usually give the right shape of the extinction, so the serious contamination of the extinction lightcurve by the intrinsic differences in the spectra is unlikely. 

The result of totally different experiment is shown in Fig.~\ref{fig:mrk335}. We simply analysed the data for a single but strongly variable object, Mrk 335. The numerous UV specrea for this object come from IUE and were taken from Dunn et al. (2006). Optical monitoring was performed in Crimean observatory, and the average optical spectra as well as the variance in U,B,V,R and I bands are given in Doroshenko et al. (2005). We used the bluest IUE spectrum and calculated the extinction curve separately for each of the remaining spectra and then we averaged the results. We then calculated the extinction curve in the optical part by taken the average plus $2\sigma$ as the reference and the average minus $2\sigma$ as the reddened spectrum. The resulting pseudo-extinction curve is shown in the right panel.

The resulting pseudo-extinction curve is not unlike other extinction curves and this may be a serious warning that contamination due to the intrinsic differences between the reddened and unreddened spectra may not be negligigle. In the case of Mrk 335 experiment the variability is at least partially {\it intrinsic}, since Sergeev et al. (2005) and Doroshenko et al. (2005) reported the measured very short time delays between the optical bands. On the other hand, part of the far UV variability may be due to the variable extinction (for examples of variable extinction see Akylas et al. 2002, Risaliti et al. 2005). The two effects are difficult to disentangle although the time-dependent modelling of the optical spectra (Czerny \& Janiuk 2006) is a good starting point.  

Mrk 335 used in this experiment is a Seyfert 1 galaxy. The SDSS sample is dominated by quasars, although the contribution of fainter objects is present. The color distribution in SDSS objects was recently studied by Bonning et al. (2006) and modeled by the theoretical accretion disk spectra. The authors argued against the importance of the dust for the color distribution although they noticed a significant discrepancy between the models and the data for objects with large maximum disk temperature (relatively narrow emission lines) although other authors indicated the increased importance of dust in Narrow Line Seyfert 1 galaxies (e.g. Kuraszkiewicz et al. 2000, Constantin \& Shields 2005). 

\section{Discussion and summary}

It is clear that the extinction curve of the circumnuclear dust does not show a strong 2175 \AA~ feature. However, we cannot determine directly from the observations whether it is as steep as the extinction in SMC or it is more grey, or shallow, in far UV since the possible systematic effects are difficult to estimate. 

Analysis of the extiction curves in other sites may, however, indicate that the extinction is not rising in UV so steeply as in SMC. The argument is the following. There are generally three reasons why the extinction curve dos not show the 2175 \AA~ feature 
\begin{itemize}
\item {low metallicity}
\item {strong radiation field}
\item {large density}.
\end{itemize}
SMC is a low metallicity galaxy (e.g. Kayser et al. 2006 and the references therein) and the featureless extinction in its bulge is most likely related to tht fact. On the other hand, also featureless extinction curve was derived for starburst galaxies and in the case of dense molecular clouds in our Galaxy, and in those cases the metallicity does not seem lower than solar but the density and the radiation field are enhanced in star forming regions. Extinction in these regions, although featureless as in SMC, does not rise so strongly in UV. Studies of AGN emission and absorption lines indicate solar or higher metallicity, even in the case of the most distant quasars. Therefore, it is more likely that physically the extinction in AGN is more similar to extinction in dense molecular clouds than to extinction in SMC bulge. 

The local gas densities in AGN dust region and in the dense molecura clouds are not widely different. The number density in Taurus dark cloud TMC-1 was estimated to be $\sim 10^4 - 10^5$ cm$^{-3}$ (Turner et al. 2005), while the constraints based on variable obscuration indicated the dust clump density $\sim 10^7$cm$^{-3}$ in a Seyfert 2 galaxy Mrk 348 (Akylas et al. 2002). The radiation field may have similar temperature $T \sim 10^5$ K in both cases, and similar intensity. 

This in turn indicates that extinction curves determined by Czerny et al. (2004) and Gaskell \& Benker et al. (2006) are likely to catch the wavelength dependence correctly. 

On the other hand, the dust properties do not only reflect the current local conditions but they also depend on the formation process. Usually the dust is supposed to form most effectively during supernova explosions (see e.g. Hirashita et al. 2005 and the references therein). In AGN, this may take place in the associated starburst, praticularly if the star formation takes place in the outer parts of an accretion disk (Collin \& Zahn 1999). The dust formation is also possible directly in the stellar wind so it is also likely to occur in the accretion disk wind, far enough from the nucleus, but the timescale for this process is perhaps too long to make it working effectively (see  e.g. Loska et al. 1993), and the dust is easily destroyed in a quasar suurounding (Weingartner et al. 2006). Further studies of the dust location (e.g. through reverberation, see e.g. Sugunuma et al. 2006), dust emission and dust absorption/scattering, if combined, can give much better insight into the dust properties.

\acknowledgements 
I thank Martin Gaskell and Moshe Elitzur more many enlightening discussions,
and Adam Dobrzycki and Asia Kuraszkiewicz for providing their data for 
PG1211+143.
Part of this work was supported by grant 
1P03D00829 of the Polish
State Committee for Scientific Research.


\end{document}